\documentclass{article} 
\usepackage{iclr2022_conference,times}

\usepackage{amsmath,amsfonts,bm}

\def\eqref#1{equation~\ref{#1}}

\def\1{\bm{1}}

\DeclareMathAlphabet{\mathsfit}{\encodingdefault}{\sfdefault}{m}{sl}
\SetMathAlphabet{\mathsfit}{bold}{\encodingdefault}{\sfdefault}{bx}{n}

\newcommand{\R}{\mathbb{R}}

\usepackage{hyperref}
\usepackage{url}

\usepackage{tabularx}
\usepackage{graphicx}
\usepackage{wrapfig}
\usepackage[export]{adjustbox}
\usepackage{booktabs}
\usepackage{float}
\usepackage{mathtools}
\usepackage{multirow}
\usepackage{multicol}

\title{Towards Fast Simulation of \\
Environmental Fluid Mechanics with \\
Multi-Scale Graph Neural Networks}

\author{
  Mario Lino \\
  Department of Aeronautics\\
  Imperial College London\\
  \texttt{mal1218@ic.ac.uk} \\
  \And
  Stathi Fotiadis \\
  Department of Bioengineering  \\
  Imperial College London \\
  \AND
  Anil A. Bharath \\
  Department of Bioengineering  \\
  Imperial College London \\
  \And
  Chris Cantwell \\
  Department of Aeronautics \\
  Imperial College London \\
}

\newcommand{\model}{{MultiScaleGNN}}
\newcolumntype{Y}{>{\centering\arraybackslash}X}

\iclrfinalcopy 
\begin{document}

\maketitle

\begin{abstract}
Numerical simulators are essential tools in the study of natural fluid-systems, but their performance often limits application in practice.
Recent machine-learning approaches have demonstrated their ability to accelerate spatio-temporal predictions, although, with only moderate accuracy in comparison.
Here we introduce \model, a novel multi-scale graph neural network model for learning to infer unsteady continuum mechanics in problems encompassing a range of length scales and complex boundary geometries.
%
%
%
We demonstrate this method on advection problems and incompressible fluid dynamics, both fundamental phenomena in oceanic and atmospheric processes.
Our results show good extrapolation to new domain geometries and parameters for long-term temporal simulations.
Simulations obtained with \model{} are between two and four orders of magnitude faster than those on which it was trained.

\end{abstract}

\section{Introduction}
Forecasting the spatio-temporal mechanics of continuous systems is a common requirement in many areas of science and engineering, including environmental fluid dynamics \citep{rubin2001environmental}.
Physical models for the transport and dispersion processes in natural fluid flows often consist on one or more partial differential equations (PDEs), whose complexity may preclude their analytic solution \citep{kim1999application}.
Numerical methods are well-established for approximating the solution of PDEs with high accuracy, but they are computationally expensive \citep{spencer}.
Deep learning techniques have been shown to be capable of accelerating physical simulations \citep{Guo2016}.  
Most of the recent work on deep learning to infer continuum physics has focused on the use of convolutional neural networks (CNNs).
In part, the success of CNNs for these problems lies in their translation invariance and locality \cite{goodfellow2016deep}, which represent strong and desirable inductive biases for continuum-mechanics models.
However, CNNs constrain input and output fields to be defined on rectangular domains represented by regular grids, which is not suitable for more complex domains.
As with traditional numerical techniques, it is desirable to be able to vary the resolution in space, devoting more effort where the physics are challenging to resolve, and less effort elsewhere.
An alternative approach to applying deep learning to geometrically and topologically complex domains is provided by graph neural networks (GNNs), which can also be designed to satisfy spatial invariance and locality \citep{battaglia2018relational,wu2020comprehensive}.

\begin{figure}[ht]
\centering
\begin{tabular}{cc}
\includegraphics[clip, height=0.30\columnwidth, trim={0mm 0mm 0mm 0mm} ]{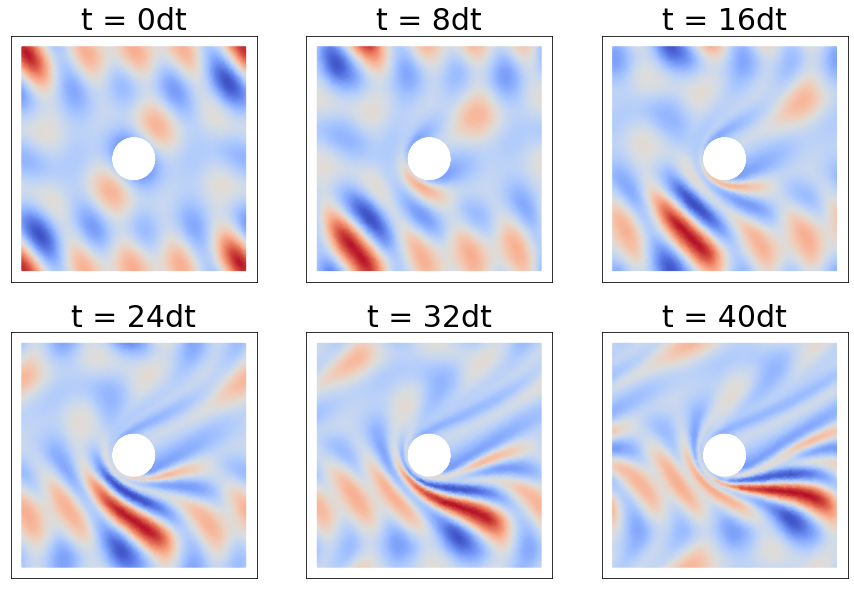}
&
\includegraphics[clip, height=0.30\columnwidth, trim={0mm 0mm 0mm 0mm}]{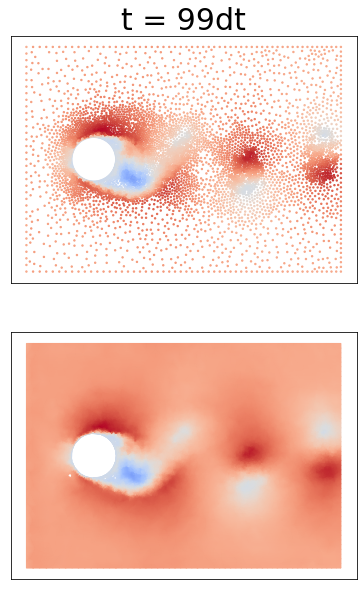} \\
{\small (a) Advection} &
{\small (b) Incompressible flow}
\end{tabular}
\caption{\small \model{} forecasts continuum dynamics. We used it to simulate (a) advection
[\href{https://imperialcollegelondon.box.com/s/a3pyd29toior6tijh4sj3yrh5x3yy95k}{\texttt{video}}]
and, (b) the incompressible flow around a circular cylinder
[\href{https://imperialcollegelondon.box.com/s/zc2e3bpgrhzrcxdb4yt0d5lzcd26npw0}{\texttt{video}}].
\label{fig:intro}
}
\end{figure}

In this paper, we describe a novel approach to applying GNNs for accurately forecasting the evolution of physical systems in complex and irregular domains.
We propose \model, a multi-scale GNN model, to forecast the spatio-temporal evolution of continuous systems discretised as unstructured sets of nodes. 
Each scale processes the information at different resolutions, enabling the network to more accurately and efficiently capture complex physical systems.
We apply \model \ to simulate advection and incompressible fluid dynamics, fundamental processes in oceanic and atmospheric research. 
Importantly, \model{} is independent of the spatial discretisation and the mean absolute error (MAE) decreases linearly as the distance between nodes is reduced, which allows for the application of adaptive re-meshing (see Figure \ref{fig:intro}b)
\model \ simulations are between two and four orders of magnitude faster than the numerical solver used for generating the ground truth datasets, becoming a potential surrogate model for fast predictions.

\section{Related work}

During the last five years, most deep neural networks (DNNs) architectures used for predicting fluid dynamics have included convolutional layers \citep{Guo2016,tompson2017accelerating,Lee2018,Kim2019,Wiewel2019,Fotiadis2020}.
These CNN-based solvers are between one and four orders of magnitude faster than numerical solvers \citep{Guo2016,Lino2020}, and some of them have shown good extrapolation to unseen domain geometries and initial conditions \citep{Thuerey2018,Lino2020}.
Recently, GNNs have been used to simulate the motion of discrete systems of solid particles \citep{battaglia2016interaction,chang2016compositional} and deformable solids and fluids discretised into Lagrangian (or \textit{free}) particles \citep{li2018learning,mrowca2018flexible,sanchez2020learning}.
Further research in this area introduced more general message-passing (MP) layers \citep{Sanchez-Gonzalez2018,Li2019a,Mrowca2018}, high-order time-integration \citep{sanchez2019hamiltonian} and hierarchical models \citep{li2018learning,mrowca2018flexible}.
To the best of our knowledge Alet, et al. (2019) \cite{alet2019graph} were the first to explore the use of GNNs to infer Eulerian mechanics by solving the Poisson PDE. However, their domains remained simple, used coarse spatial discretisations and did not explore the generalisation of their model.
More closely related to our work, \cite{pfaff2020learning} proposed a mesh-based GNN to simulate continuum mechanics, although they did not consider the use of MP at multiple scales of resolution.
\cite{li2020multipole} and \cite{liu2021multi} used multi-resolution GNNs to infer steady solutions, but their pooling remained simple and did not explore extrapolation to unseen domain geometries or physical parameters.
\cite{li2020fourier} also considered multi-scale neural PDE modelling, with the mayor drawback that the spatial discretisations are constrained to uniform grids.

\section{Model}
\subsection{Model definition}

\begin{figure}[ht]
\centering
\includegraphics[clip,width=0.9\columnwidth, trim={0mm 0mm 0mm 0mm}]{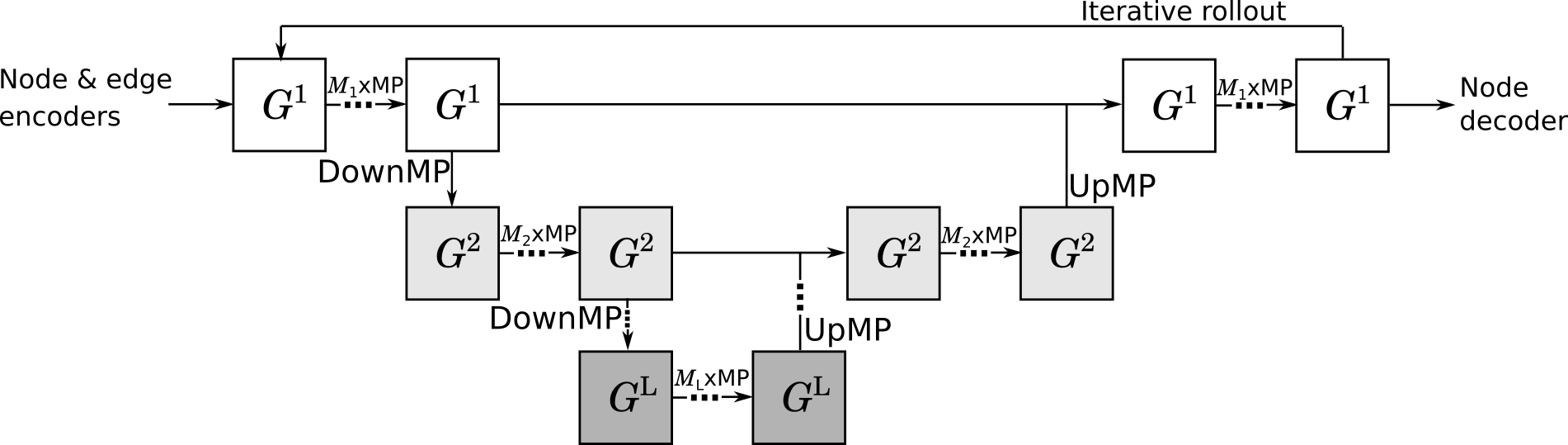}
\caption{\small \model{} architecture. $G^1$ is the high-resolution graph, $G^2$ and $G^L$ are low-resolution graphs.
}\label{fig:net}
\end{figure}

For a PDE $\frac{\partial \mathbf{u}}{\partial t}=\mathcal{F}(\mathbf{u})$ on a spatial domain $\mathcal{D} \subset \R^2$, \model \ infers the temporal evolution of $\mathbf{u}(t,\mathbf{x})$ at a finite set of nodes $V^1$, with coordinates $\mathbf{x}^1_i  \in \mathcal{D}$.
Given an input field $\mathbf{u}(t_0, \mathbf{x}_{V^1})$, at time $t=t_0$ and at the $V^1$ nodes, a single evaluation of
\model{} returns $\mathbf{u}(t_0+dt, \mathbf{x}_{V^1})$, where $dt$ is a fixed time-step size.
Each time-step is performed by applying MP layers in $L$ graphs and between them, as illustrated in Figure \ref{fig:net}.
The high-resolution graph $G^1$ consists of the set of nodes $V^1$
and a set of directed edges $E^1$ connecting these nodes.
In a complete graph, there exist $|V^1|(|V^1|-1)$ edges, this would result in a high computational cost to MP.
Instead, \model{} connects each node in $V^1$ (receiver node) to its $k$-closest nodes ($k$ sender nodes) using a $k$-nearest neighbours ($k$-NN) algorithm.
This guarantees that, in MP layers, each node is receiving information from exactly $k$ nodes, in contrast to \cite{pfaff2020learning} and \cite{Sanchez-Gonzalez2018}. This uniformity can ease the learning and improve the accuracy of the predictions.
The attributes assigned to each node $\mathbf{v}^1_i$ are the concatenation of $\mathbf{u}(t_0, \mathbf{x}_{i})$, $\mathbf{p}_i$ and $\Omega_i$, where
$\mathbf{p}_i$ is a set of physical parameters at $\mathbf{x}_i$ (such as the Reynolds number in fluid dynamics) and $\Omega_i = 1$ for nodes located on Dirichlet boundaries and zero elsewhere.
Each edge attribute $\mathbf{e}^1_k$ is assigned the relative position between sender node $s_k$ and receiver node $r_k$.

Node attributes and edge attributes are encoded through two independent multi-layer perceptrons (MLPs).
A MP layer applied to $G^1$ propagates the nodal and edge information only locally between adjacent nodes.
Nevertheless, most continuous physical-systems require this propagation at larger scales, or even globally.
To handle this, \model{} processes the information at $L$ scales, creating a graph for each level and propagating information between them in each pass.
The lower-resolution graphs ($G^2, G^3, \dots, G^L$; with $|V_1| > |V_2| > \dots > |V_L|$) possess fewer nodes and edges, and hence, a single MP layer can propagate features over longer distances more efficiently.
As depicted in Figure \ref{fig:net}, the information is first diffused and processed in the high-resolution graph $G^1$ through $M_1$ MP layers. It is then passed to $G^2$ through a downward MP (DownMP) layer.
In $G^2$ the attributes are again processed through $M_2$ MP layers and a DownMP layer to $G^3$.
This process is repeated $L-1$ times. The lowest resolution attributes (stored in $G^L$) are then passed back to the scale immediately above through an upward message passing (UpMP) layer.
Attributes are again successively passed through $M_l$ MP layers at scale $l$ and an UpMP layer from scale $l$ to scale $l-1$ until the information is ultimately processed in $G^1$.
Finally, a MLP decodes the nodal information to return the predicted field at time $t_0 + dt$ at the $V^1$ nodes.
To apply MP in the $L$ graphs, \model{} uses the MP layer introduced by \cite{Sanchez-Gonzalez2018} and \cite{battaglia2018relational}, with the mean as the aggregation function.

\subsection{Multi-scale graphs}
\label{sec:multiscale_graphs}
Each graph $G^l=(V^l, E^l)$ with $l = 2, 3, \dots, L$; is obtained from graph $G^{l-1}$ by first dividing $\mathcal{D}$ into a regular grid with cell size $d_x^l \times d_y^l$ (Figure \ref{fig:pool}a exemplifies this for $l=2$).
For each cell, provided that there is at least one node from $V^{l-1}$ in it, a node is added to $V^l$.
The nodes from $V^{l-1}$ in a given cell $i$ and the node from $V^{l}$ on the same cell are denoted as child nodes, $Ch(i) \subset V^{l-1}$, and parent node, $i \in V^l$, respectively.
The coordinates of each parent node is the mean position of its children nodes.
Each edge $k \in E^{l}$ connects sender node $s_k \in V^{l}$ to receiver node $r_k \in V^l$, provided that there exists at least one edge from $Ch(s_k)$ to $Ch(r_k)$ in $E^{l-1}$.
Edge $k$ is assigned the mean edge attribute of the set of edges going from $Ch(s_k)$ to $Ch(r_k)$.

\begin{small}
\begin{figure}[ht]
\centering
\begin{tabular}{cc}
\includegraphics[clip, height=0.23\columnwidth, trim={0mm, -50mm, 0mm, 0mm}]{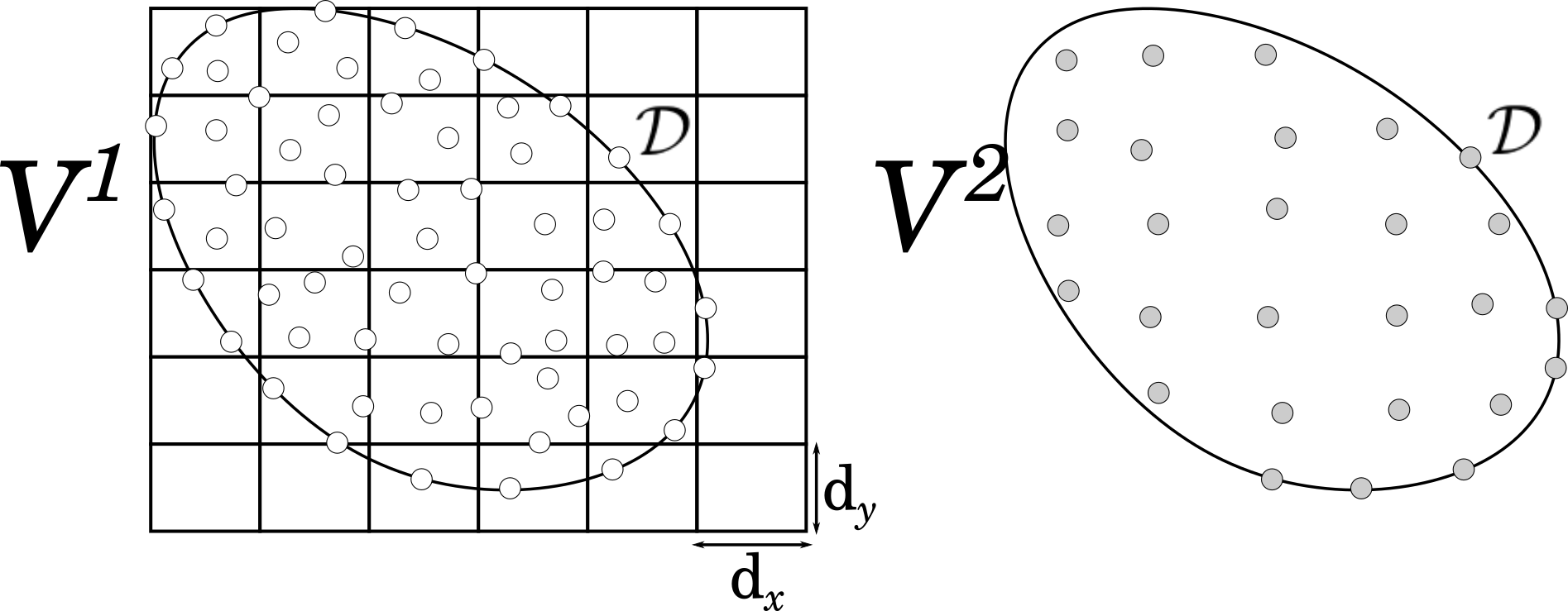} &
\includegraphics[height=0.23\columnwidth]{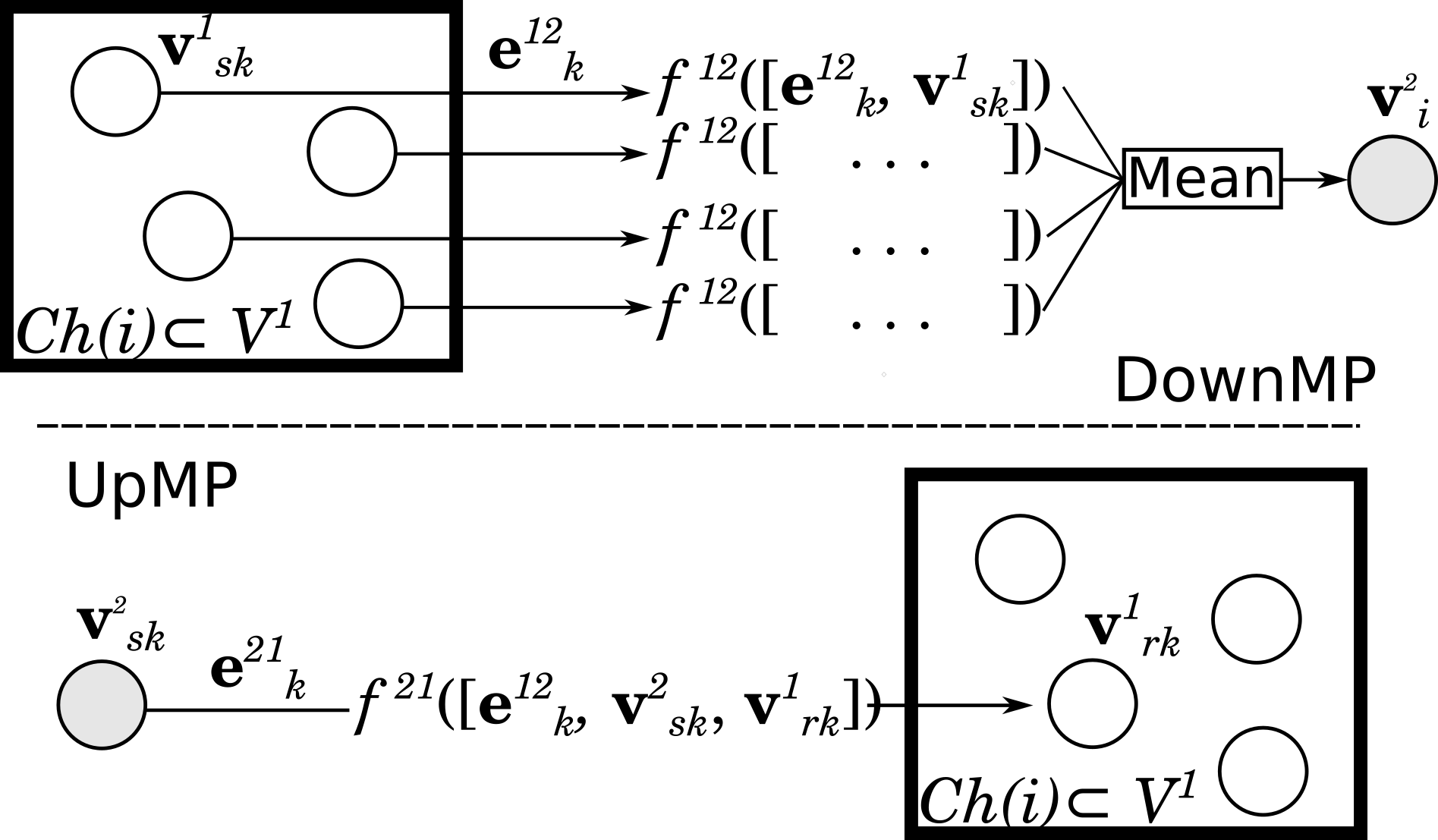} \\
{\small (a) Building $V^2$ from $V^1$} &
{\small (b) DownMP and UpMP layers}
\end{tabular}
\caption{ \small (a) $V^2$ is obtained from $V^1$ by partitioning $\mathcal{D}$ using a grid with cell size $d_x^2 \times d_y^2$. (b) DownMP and UpMP diagrams.
\label{fig:pool}
}
\end{figure}
\end{small}

\paragraph{Downward message-passing (DownMP).}
To perform MP from $G^{l-1}$ to $G^l$ (see Figure \ref{fig:pool}b), a set of directed edges, $E^{l-1,l}$, is created.
Each edge $k \in E^{l-1,l}$ connects node $s_k \in V^{l-1}$ to its parent node $r_k \in V^{l}$, with edge attributes assigned as the relative position between child and parent nodes.
A DownMP layer applies a common edge-update function, $f^{l-1,l}$, to edge $k$ and node $s_k$.
It then assigns to each node attribute $\mathbf{v}^l_i$ the mean updated attribute of all the edges in $E^{l-1,l}$ arriving to node $i$, i.e.,

\begin{equation}
    \mathbf{v}^l_i = \frac{1}{|Ch(i)|} \sum_{k: r_k=i} f^{l-1, l}([\mathbf{e}^{l-1, l}_k,  \mathbf{v}^{l-1}_{s_k}]), \ \ \ \forall i \in 1, \ldots, |V^{l}|.
    \label{eq:pool}
\end{equation}

\paragraph{Upward message-passing (UpMP).}
To pass and process the node attributes from $G^{l+1}$ to $G^{l}$, \model \ defines a set of directed edges, $E^{l+1, l}$.
These edges are the same as in $E^{l,l+1}$, but with opposite direction.
An UpMP layer applies a common edge-update function, $f^{l+1,l}$, to each edge $k \in E^{l+1,l}$ and both its sender (at scale $l+1$) and receiver (at scale $l$) nodes, directly updating the node attributes in $G^{l}$, i.e.,
\begin{equation}
    \mathbf{v}^{l}_{r_k} = f^{l+1,l}([\mathbf{e}^{l+1,l}_k, \mathbf{v}^{l+1}_{s_k}, \mathbf{v}^{l}_{r_k}]), \ \ \ \forall k \in E^{l+1,l}.
    \label{eq:unpool}
\end{equation}
UpMP layers leave the edge attributes of $G^{l}$ unaltered.
To model functions $f^{l-1,l}$ and $f^{l+1,l}$, we use MLPs.

\section{Training datasets}
Datasets \texttt{AdvBox} and \texttt{AdvInBox}, both used simultaneously for training, contain simulations of a scalar field advected under a uniform velocity field on a square domain ($[0,1]\times[0,1]$) and a rectangular domain ($[0,1]\times[0,0.5]$) respectively.
\texttt{AdvBox} domains have periodic conditions on all four boundaries, whereas \texttt{AdvInBox} domains have upper and lower periodic boundaries, a prescribed Dirichlet condition on the left boundary, and a zero-Neumann condition on the right boundary.
The initial states at $t_0$ are derived from two-dimensional truncated Fourier series with random coefficients and a random number of terms.
For advection models, $\mathbf{u}(t,\mathbf{x}_i) \in \R$ is the advected field and $\mathbf{p}_i \in \R^2$ are the two components of the velocity field at $\mathbf{x}_i$.

Dataset \texttt{NS} contains simulations of the periodic vortex shedding around a circular cylinder at Reynolds numbers between 500 and 1000.
The upper and lower boundaries are periodic, and the vertical distance between cylinders is randomly sampled between 4 and 6.  
Each domain is discretised into approximately 7000 nodes.
For flow models, $\mathbf{u}(t,\mathbf{x}_i) \in \R^3$ contains the velocity and pressure fields and $\mathbf{p}_i \in \R$ is the Reynolds number.
Further details of the training and testing datasets are included in Appendix \ref{sec:data_detail}.

\section{Results and discussion}

\begin{wrapfigure}{R}{0.35\columnwidth}
\begin{center}
\includegraphics[clip, width=0.35\columnwidth, trim={0mm, 0mm, 0mm, 0mm}]{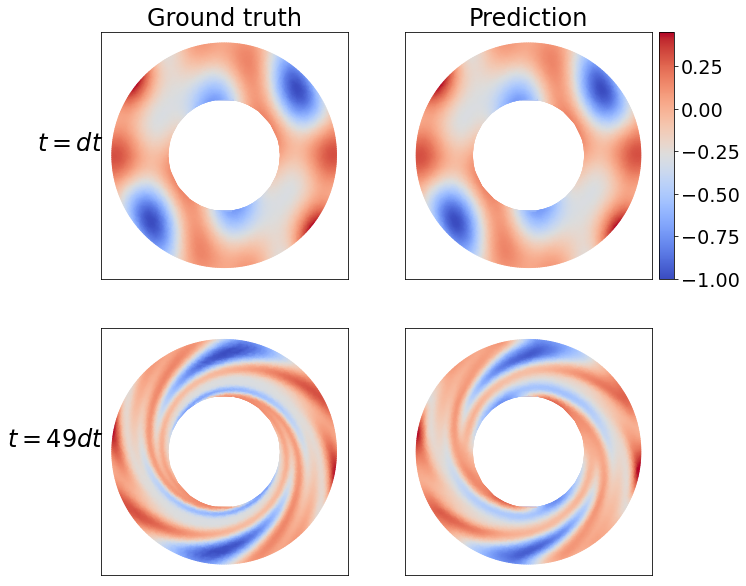}
\end{center}
\caption{ \small Advection of a scalar field in a Taylor-Couette flow [\href{https://imperialcollegelondon.box.com/s/jvemqgeffwgitall02enk4wo4237n8wv}{\texttt{video}}].
\label{fig:taylor_splines}
}
\end{wrapfigure}

We first consider the \model{} model trained to infer advection.
Despite \model{} being trained on square and rectangular domains and uniform velocity fields, it generalises to complex domains and non-uniform velocity fields.
As an example of a closed domain, we consider the Taylor-Couette flow in Figure \ref{fig:taylor_splines}, where after 49 time-steps a \model{} with $L=1$ maintains high accuracy in transporting both the lower and the higher frequencies generated due to the shear flow.
We also evaluated the predictions of \model{} on open domains containing obstacles of different shapes (circles, squares, ellipses and closed spline curves).
Figure \ref{fig:intro} shows the predictions obtained with a \model{} with $L=2$ for a field advected around a circular cylinder and upper-lower and left-right periodic boundaries.
\model{} was also trained to simulate unsteady incompressible fluid dynamics in a range of Reynolds numbers between 500 and 1000.
It shows very good interpolation to unseen Reynolds numbers within this range.
For instance, Figure \ref{fig:intro}b shows the horizontal velocity field predicted by \model{} with $L=4$ after 99 time-steps for $Re=700$.
The MAE increases for Reynolds numbers lower than 400 and higher than 1200, but the predictions remain visually realistic in the range 300 to 1500, indicating that the network has learnt some fundamental aspects of the fluid dynamics.
\footnote{Animations with the ground truth and best-model predictions can be found \href{https://imperialcollegelondon.box.com/s/f6eqb25rt14mhacaysqn436g7bup3onz}{\textit{here}}.}

We evaluate the accuracy of \model{} with $L=1,2,3$ and $4$; the architectural details of each model are included in Appendix \ref{sec:model_detail}.
Tables \ref{table:mae_adv_scales} and \ref{table:mae_ns_scales} collect the MAE for the last time-point and the mean of all the time-points on the testing datasets.
Incompressible fluids have a global behaviour, and the addition of coarser layers helps the network to learn this characteristic and achieve significantly lower errors.
Hence, for the N-S testing dataset there is a clear benefit from increasing the number of scales.
In contrast, for the advection datasets, the lowest MAE values are obtained for $L=1$ or 2, since in advection the information is propagated only locally.
As a comparison to \cite{pfaff2020learning}, a GNN with 16 sequential MP-layers (GN-Blocks) results in a MAE of $5.852 \times 10^{-2}$ on our \texttt{NSMidRe} dataset; whereas \model{} with the same number and type of MP layers, but distributed among 3 scales, results in a lower MAE of $3.081 \times 10^{-2}$.
A comparison of our coarsening/pooling algorithm to \cite{liu2021multi} is included in Appendix \ref{sec:coarsening}.

\begin{table}[ht]
\caption{\small MAE $\times 10^{-2}$ on the advection testing datasets for \model \ models with $L = 1, 2, 3, 4$} \label{table:mae_adv_scales}
\begin{center}
\footnotesize
\begin{tabularx}{\textwidth}{lYY|YY|YY|YY}
\toprule
\multirow{2}{*}{Datasets} & \multicolumn{2}{c}{$L=1$} & \multicolumn{2}{c}{$L=2$} & \multicolumn{2}{c}{$L=3$} & \multicolumn{2}{c}{$L=4$} \\

                  &
Step 49 & All &
Step 49 & All &
Step 49 & All &
Step 49 & All \\

\midrule

\texttt{AdvTaylor} & \textbf{6.940} & \textbf{3.195} & 7.914 & 3.676 & 8.037 & 3.739 & 8.790 & 4.050 \\

\texttt{AdvCircle} & 4.057 & 2.112 & \textbf{3.690} & \textbf{1.817} & 3.911 & 1.916 & 3.827 & 1.995 \\

\texttt{AdvCircleAng} & \textbf{3.870} & 2.030 & 3.962 & \textbf{1.890} & 4.300 & 2.074 & 4.428 & 2.249 \\

\texttt{AdvSquare} & 4.250 & 2.178 & \textbf{4.175} & \textbf{1.991} & 4.420 & 2.107 & 4.279 & 2.141 \\

\texttt{AdvEllipseH} & 4.462 & 2.241 & \textbf{4.328} & \textbf{2.017} & 4.567 & 2.136 & 4.449 & 2.179
 \\

\texttt{AdvEllipseV} & 4.265 & 2.236 & \textbf{4.132} & \textbf{2.014} & 4.297 & 2.100 & 4.207 & 2.22 \\

\texttt{AdvSplines} & 4.484 & 2.293 & \textbf{4.426} & \textbf{2.088} & 4.636 & 2.199 & 4.58 & 2.300 \\

\texttt{AdvInCir} & \textbf{11.7} & \textbf{7.101} & 25.369 & 18.548 & 27.981 & 22.626 & 27.379 & 22.034 \\

\bottomrule
\end{tabularx}
\end{center}
\label{tab:adv_scales}
\end{table}

\begin{table}[ht]
\caption{\small MAE $\times 10^{-2}$ on the N-S testing datasets for \model \ models with $L = 1, 2, 3, 4$} \label{table:mae_ns_scales}
\begin{center}
\footnotesize
\begin{tabularx}{\textwidth}{lYY|YY|YY|YY} 
\toprule
\multirow{2}{*}{Datasets} & \multicolumn{2}{c}{$L=1$} & \multicolumn{2}{c}{$L=2$} & \multicolumn{2}{c}{$L=3$} & \multicolumn{2}{c}{$L=4$} \\

                  &
Step 99 & All &
Step 99 & All &
Step 99 & All &
Step 99 & All \\

\midrule

\texttt{NSMidRe} & 9.765 & 6.108 & 4.759 & 3.663 & 3.851 & 3.081 & \textbf{3.456} & \textbf{2.825}\\

\texttt{NSLowRe} & 9.43 & 7.327 & 12.346 & 8.211 & 11.707 & 8.203 & \textbf{7.338} & \textbf{5.532} \\

\texttt{NSHighRe} & 9.487 & 10.598 & 7.879 & 9.002 & 6.98 & 7.096  & \textbf{5.826} & \textbf{5.871} \\

\bottomrule
\end{tabularx}
\end{center}
\end{table}

\section{Conclusion}
\model \ is a novel multi-scale GNN model for inferring mechanics on continuous systems discretised into unstructured sets of nodes.
Unstructured discretisations allow complex domains to be accurately represented and the node count to be adjusted over space. Multiple coarser levels allow high and low-resolution mechanics to be efficiently captured.
In global and local problems, such as incompressible fluid dynamics, 
the coarser graphs are particularly advantageous, since they enable global characteristics to be learnt.
\model{} interpolates to unseen spatial discretisations of the physical domains, allowing it to adopt efficient discretisations and to dynamically and locally modify them to further improve the accuracy.
\model{} also generalises to advection on complex domains and velocity fields and it interpolates and extrapolates to unseen Reynolds numbers in fluid dynamics.
Inference is between two and four orders of magnitude faster than with the high-order solver used for generating the training datasets.
This work is a significant advancement in the design of flexible, accurate and efficient neural simulators for fluid dynamics and in general for continuum mechanics.

\bibliography{iclr2022_conference}
\bibliographystyle{iclr2022_conference}

\appendix

\section{Datasets details} \label{sec:data_detail}

\subsection{Advection datasets} \label{sec:adv_datasets}

We solved the two-dimensional advection equation using Nektar++, an spectral/hp element solver  \citep{spencer,nektar}.
As initial condition, $\varphi_0$, we take an scalar field derived from a two-dimensional Fourier series with $M \times N$ random coefficients, specifically
\begin{gather}
    \varphi_0 = \sum_{m=0}^M \sum_{n=0}^N c_{m,n} \varphi_{m,n} \cdot \exp \Big(-2 (x-x_c)^2 -2 (y-y_c)^2 \Big), \label{eq:ic} \\
    \text{with} \ \ \ \varphi_{m,n} = \Re \bigg \{ \exp\Big(i 2 \pi (mx+ny) \Big)  \bigg \}. \label{eq:phi_mn}
\end{gather}
Coefficients $c_{m,n}$ are sampled from a uniform distribution between 0 and 1, and integers $M$ and $N$ are randomly selected between 3 and 8.
In equation (\ref{eq:phi_mn}), $x_c$ and $y_c$ are the coordinates of the centre of the domain. 
The initial field $\varphi_0$ is scaled to have a maximum equal to $1$ and a minimum equal to $-1$.
We created training and testing datasets containing advection simulations with 50 time-points each, equispaced by a time-step size $dt = 0.03$. 
A summary of these datasets can be found in Table \ref{table:adv_datasets}.

\paragraph{Training datasets.}
We generated two training datasets: \texttt{AdvBox} with 1500 simulations and \texttt{AdvInBox} with 3000 simulations.
In these datasets we impose a uniform velocity fields with random values for $u$ and $v$, but constrained to $u^2+v^2 \leq 1$.
In dataset \texttt{AdvBox} the domain is a square ($\mathbf{x} \in [0,1]\times[0,1]$) with periodicity in $x$ and $y$.
In dataset \texttt{AdvInBox} the domain is a rectangle ($\mathbf{x} \in [0,1]\times[0,0.5]$) with periodicity in $y$, a Dirichlet condition on the left boundary and a homogeneous Neumann condition on the right boundary -- as an additional constraint, $u \geq 0$.
During training, a new set of nodes $V^1$ is selected at the beginning of every iteration.
The node count was varied smoothly across the different regions of the domains.
The sets of node were created with Gmsh, a finite-element mesher. 
The \textit{element size} parameter was set to 0.012 in the corners and the centre of the training domains, and set to $\sqrt{10}$ or $1/\sqrt{10}$ times that value at one random control point on each boundary. 
The mean number of nodes in $|V_1|$ for \texttt{AdvBox} and \texttt{AdvInBox} predictions are 9802 and 5009 respectively. 

\begin{figure}[ht]
\begin{small}
    \centering
    \begin{adjustbox}{minipage=\linewidth,scale=0.9}
    \centering
    \begin{tabular}{cccccc}
        \raisebox{-\height}{(a)} &
        \raisebox{-\height}{
        \includegraphics[width=0.17\textwidth]{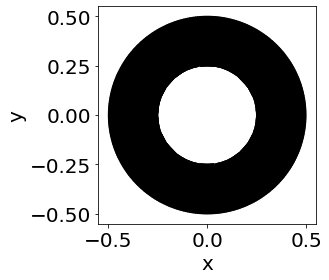}}
        &
        \raisebox{-\height}{(b)} &
        \raisebox{-\height}{
        \includegraphics[width=0.17\textwidth]{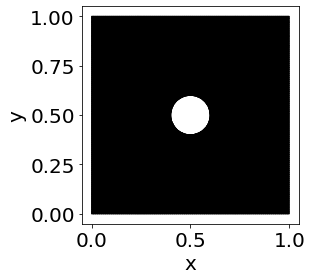}}
        &
        \raisebox{-\height}{(c)} &
        \raisebox{-\height}{
        \includegraphics[width=0.17\textwidth]{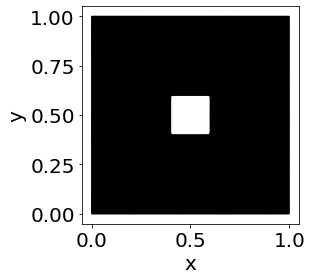}}
        \\
        \raisebox{-\height}{(d)} &
        \raisebox{-\height}{
        \includegraphics[width=0.17\textwidth]{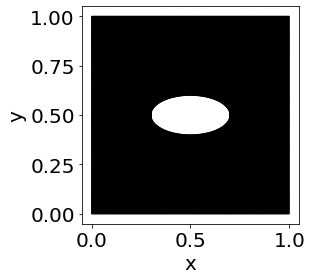}}
        &
        \raisebox{-\height}{(e)} &
        \raisebox{-\height}{
        \includegraphics[width=0.17\textwidth]{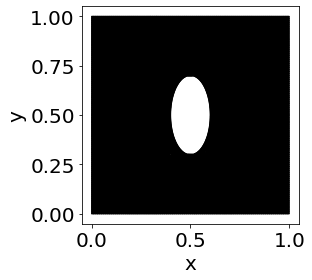}}
        &
        \raisebox{-\height}{(f)} &
        \raisebox{-\height}{
        \includegraphics[width=0.17\textwidth]{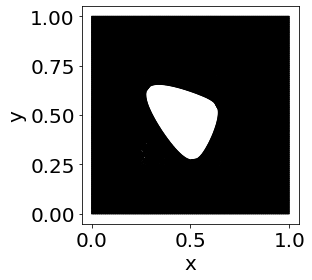}}
    \end{tabular}
    \end{adjustbox}
    \caption{\small Physical domains (black areas) on our testing datasets.}
    \label{fig:geo}
\end{small}
\end{figure}

\paragraph{Testing datasets.} 
We generated eight testing datasets, each of them containing 200 simulations.
These datasets consider advection on more complex open and closed domains with non-uniform velocity fields.
The domains employed are represented in Figure \ref{fig:geo}, and the testing datasets are listed in Table \ref{table:adv_datasets}.
The velocity fields were obtained from the steady incompressible Navier-Stokes (N-S) equations with $Re=1$.
In dataset \texttt{AdvTaylor} the inner and outer walls spin at a velocity randomly sampled between $-1$ and $1$.
In datasets \texttt{AdvCircle}, \texttt{AdvSquare}, \texttt{AdvEllipseH}, \texttt{AdvEllipseV} and \texttt{AdvSplines} there is periodicity along $x$ and $y$, and a horizontal flow rate between $0.2$ and $0.75$ is imposed.
The obstacles inside the domains on the \texttt{AdvSplines} dataset are made of closed spline curves defined from six random points.
Dataset \texttt{AdvCircleAng} is similar to \texttt{AdvCircle}, but the flow rate forms an angle between $-45 \deg$ and $45 \deg$ with the $x$ axis.
The domain in dataset \texttt{AdvInCir} has periodicity along $y$, a Dirichlet condition on the left boundary (with $0.2 \leq u^2 + v^2  \leq 0.75$ and $-45 \deg  \leq \arctan(v/u)  \leq 45 \deg$), and a homogeneous Neumann condition on the right boundary.
The set of nodes $V^1$ were generated using Gmsh with an element size equal to 0.005 on the walls of the obstacles and 0.01 on the remaining boudaries.

\begin{table}[ht]
\caption{\small Advection training and testing datasets}
\label{table:adv_datasets}
\begin{center}
\scriptsize
\begin{tabularx}{\textwidth}{lcccc} 
\toprule
Dataset                 & Flow type                      & Domain  & \#Nodes               & Train/Test   \\ 
\midrule
\texttt{AdvBox}      & Open, periodic in $x$ and $y$ & $[0,1] \times [0,1]$     & 9601-10003            & Training \\
\texttt{AdvInBox}      & Open, periodic in $y$ & $[0,1] \times [0,0.5]$     & 4894-5135            & Training \\
\texttt{AdvTaylor}      & Closed, Taylor-Couette flow             & Figure \ref{fig:geo}a    & 7207           & Testing \\
\texttt{AdvCircle}      & Open, periodic in $x$ and $y$ & Figure \ref{fig:geo}b     & 19862            & Testing \\
\texttt{AdvCircleAng}   & Open, periodic in $x$ and $y$ & Figure \ref{fig:geo}b     & 19862           & Testing \\
\texttt{AdvSquare}      & Open, periodic in $x$ and $y$ & Figure \ref{fig:geo}c     & 19956            & Testing \\
\texttt{AdvEllipseH}    & Open, periodic in $x$ and $y$ & Figure \ref{fig:geo}d     & 20210            & Testing \\
\texttt{AdvEllipseV}    & Open, periodic in $x$ and $y$ & Figure \ref{fig:geo}e     & 20221           & Testing \\
\texttt{AdvSplines}     & Open, periodic in $x$ and $y$ & Figure \ref{fig:geo}f     & 19316-20389            & Testing \\
\texttt{AdvIncir}       & Open, periodic in $y$         & Figure \ref{fig:geo}b     & 19862           & Testing \\
\bottomrule
\end{tabularx}
\end{center}
\end{table}

\subsection{Incompressible fluid dynamics datasets} \label{sec:ns_datasets}
We solved the two-dimensional incompressible Navier-Stokes equation using the high-order solver Nektar++.
We consider the flow around an infinite vertical array of circular cylinders, with diameter $D=1$, equispaced a distance $H$ randomly sampled between $4D$ and $6D$.
The width of the domain is $7D$ and the cylinders axis is at $1.5D$ from the left boundary.
The left boundary is an inlet with $u = 1$, $v=0$ and $\partial p/ \partial x = 0$; the right boundary is an outlet with $\partial u/ \partial x = 0$, $\partial v/ \partial x=0$ and $p = 0$; and, the cylinder walls have a no-slip condition. 
In our simulations, we select $Re$ values that yield solutions in the laminar vortex-shedding regime, and we only include the periodic stage to our datasets.
The sets of nodes $V^1$ employed for each simulation were created using Gmsh placing more nodes around the cylinders walls.
The mean number of nodes in these sets is 7143.
Each simulation contains $100$ time-points equispaced by a time-step size $dt=0.1$.
The training and testing datasets are listed in Table \ref{table:ns_datasets}.

\begin{table}
\parbox{.55\linewidth}{
\caption{\small Incompressible flow training and testing datasets}
\label{table:ns_datasets}
\center
\small
\begin{tabular}{lccc} 
\toprule
Dataset                 & $Re$ & \#Simulations    & Train/Test   \\ 
\midrule
\texttt{NS}      & 500-1000 & 1000  & Training \\
\texttt{NSMidRe} & 500-1000 & 250 & Testing \\
\texttt{NSLowRe} & 100-500 & 250 & Testing \\
\texttt{NSMidRe} & 1000-1500 & 250 & Testing \\
\bottomrule
\end{tabular}
}
\parbox{.45\linewidth}{
\caption{\small Cell sizes for coarsening to levels 2, 3 and 4}
\label{table:hyper}
\center
\small
\begin{tabular}{ccc} 
\toprule
Cell size & Advection & Fluid dynamics   \\ 
\midrule
$d_x^2, d_y^2$ & 0.02 & 0.15 \\
$d_x^3, d_y^3$ & 0.04 & 0.30 \\
$d_x^4, d_y^4$ & 0.08 & 0.60 \\
\bottomrule
\end{tabular}
}
\end{table}

\section{Model details} \label{sec:model_detail}

\paragraph{Hyper-parameters choice.}
The number of edges going to each node was set to $k=6$, and the number of layers in each MLP (encoder, decoder and edge and node update functions) was three, with 128 neurons per hidden layer.
All MLPs (except the decoder) use SELU activation functions \citep{klambauer2017self}, and, batch normalisation \citep{ba2016layer}.
The grid sizes used for generating the coarser graphs are collected in Table \ref{table:hyper}.
The number of MP layers we used at each scale are listed in the Table \ref{table:hyper2}.

\begin{table}[ht]
\centering
\caption{\small Number of MP layers at each scale for $L=1,2,3$ and 4}
\label{table:hyper2}
\begin{center}
\small
\begin{tabular}{c|cc} 
\toprule
$L$                 & Advection & Fluid dynamics   \\ 
\midrule
$L=1$   & $M_1=4$  & $M_1=8$  \\
$L=2$   & $M_1=2, M_2=4$  & $M_1=4, M_2 = 4$  \\
$L=3$   & $M_1=2, M_2=2, M_3 = 4 $  & $M_1=4, M_2 = 2, M_3 = 4$  \\
$L=4$   & $M_1=2, M_2=2, M_3 = 2, M_4=4 $  & $M_1=4, M_2 = 2, M_3 = 2, M_4=4$  \\
\bottomrule
\end{tabular}
\end{center}
\end{table}

\paragraph{Training details.}
We trained \model \ models on a internal cluster using 4 CPUs, 86GB of memory, and a RTX6000 GPU with 24GB.
We fed 8 graphs per batch.
First, each training iteration predicted a single time-point, and, every time the training loss decreased below a threshold (0.01 for advection and 0.005 for fluid dynamics), we increased the number of iterative time-steps by one, up to a limit of 10.
We used the loss function given by 
\begin{align}
    \mathcal{L} = \mathrm{MSE}&\Big(\hat{\mathbf{u}}(t, \mathbf{x}^1_{V^1}),\mathbf{u}(t, \mathbf{x}^1_{V^1}) \Big) \nonumber\\
    &+ \lambda_d\, \mathrm{MAE}\Big(\hat{\mathbf{u}}(t, {\mathbf{x}^1_{V^1} \in \partial_D\mathcal{D}}),\mathbf{u}(t, \mathbf{x}^1_{V^1} \in \partial_D\mathcal{D}) \Big) \nonumber\\
    &+ \frac{\lambda_e}{|E^1|} \sum_{\forall k \in E^1} \mathrm{MSE}\bigg(\frac{\hat{\mathbf{u}}(t, \mathbf{x}^1_{r_k}) - \hat{\mathbf{u}}(t, \mathbf{x}^1_{s_k})}{||\mathbf{e}_k||_2}, \frac{\mathbf{u}(t, \mathbf{x}^1_{r_k}) - \mathbf{u}(t, \mathbf{x}^1_{s_k})}{||\mathbf{e}_k||_2}\bigg).
\label{eq:loss}
\end{align}
with $\lambda_d=0.25$, and, $\lambda_e=0.5$ for advection and $\lambda_e=0$ for fluid dynamics.
The initial time-point was randomly selected for each prediction, and, we added to the initial field noise following a uniform distribution between -0.01 and 0.01. 
After each time-step, the models' weights were updated using the Adam optimiser with its standard parameters \cite{kingma2014adam}.
The learning rate was set to $10^{-4}$ and multiplied by 0.5 when the training loss did not decrease after six consecutive epochs, also, we applied gradient clipping to keep the Frobenius norm of the weights' gradients below or equal to one.

\section{Coarsening comparison}
\label{sec:coarsening}
An important question in the design of \model{} was how to coarsen $V_l$ to $V_{l+1}$.
In the field of numerical simulations several coarsening algorithms have been developed to guarantee the stability and accuracy of the simulations.
\cite{liu2021multi} and \cite{belbute2020combining} used coarsening techniques from numerical simulations and then interpolated the nodes attributes into the coarsened set of nodes.
We found that our \textit{cell-grid} coarsening and learnt MP from the high to the low-resolution graph (see section \ref{sec:multiscale_graphs}) performs significantly better than such coarsening-interpolation approach.
Table \ref{table:mae_coarsenings} shows the MAE on the incompressible flow testing datasets for \model{} models using both pooling strategies and a random coarsening, which randomly drops nodes to match the same $|V_{l}|/|V_{l+1}|$ ratio.
In contrast to coarsening algorithms adopted from numerical simulations, cell-grid coarsening does not preserve the spatial node density from the original discretisation, but it results into a pseudo-structured discretisation. 
This discretisation may help to spread the node information more efficiently on all directions and regions and to ease the learning of the parameters of the MP layers at low resolution scales.
Also, instead of a vanilla interpolation, in the DownMP the node information is passed to the low-resolution graph using a learnt MP layer, and the edge information is also passed to the edges of the low-resolution graph.

\begin{table}[ht]
\centering
\caption{\small MAE $\times 10^{-2}$ on the N-S testing datasets for $L = 1, 2, 3, 4$ with three different coarsening algorithms} \label{table:mae_coarsenings}
\begin{center}
\scriptsize
\begin{tabularx}{\textwidth}{ll|YY|YY|YY} 
\toprule
\multirow{2}{*}{Datasets} & \multirow{2}{*}{$L$} & \multicolumn{2}{c}{Cell grid} & \multicolumn{2}{c}{Guillard's} & \multicolumn{2}{c}{Random}\\

              &
              &
Step 99 & All &
Step 99 & All &
Step 99 & All \\

\midrule

\multirow{3}{*}{\texttt{NSMidRe}} & 2 & \textbf{4.759} & \textbf{3.663} & 6.756 & 4.393 & 7.191 & 4.741 \\
                                  & 3 & \textbf{3.851} & \textbf{3.081} & 5.735 & 3.952 & 6.431 & 4.606 \\
                                  & 4 & \textbf{3.456} & \textbf{2.825} & 4.494 & 3.330 & 6.337 & 4.428 \\

\midrule

\multirow{3}{*}{\texttt{NSLowRe}} & 2 & 12.346 & 8.211 &  \textbf{9.053} & \textbf{6.378} & 10.736 & 7.255\\
                                  & 3 & 11.707 & 8.203 &  9.099 & 6.946 & \textbf{8.977} & \textbf{6.672} \\
                                  & 4 &  \textbf{7.338} & \textbf{5.532} & 14.358 & 9.164 & 10.335 & 7.521\\

\midrule
                         
\multirow{3}{*}{\texttt{NSHighRe}} & 2 & \textbf{7.879} & 9.002 & 10.776 & 7.608 & 11.024 & \textbf{7.842}\\
                                   & 3 & \textbf{6.980} & \textbf{7.096} & 10.331 & 7.556 & 9.631 & 7.28\\
                                   & 4 & \textbf{5.826} & \textbf{5.871} & 11.222 & 7.837 & 9.572 & 7.202\\

\bottomrule
\end{tabularx}
\end{center}
\end{table}


\end{document}